\documentclass[conference]{IEEEtran}
\usepackage{graphicx}
\usepackage{amsmath,amsfonts,amssymb}
\usepackage{graphicx}
\usepackage{cite}
\usepackage{bm}
\usepackage{comment}
\usepackage{algorithmic}
\usepackage{algorithm}
\usepackage{array}
\usepackage[caption=false,font=normalsize,labelfont=sf,textfont=sf]{subfig}
\usepackage{textcomp}
\usepackage{stfloats}
\usepackage{url}
\usepackage{verbatim}
\usepackage{graphicx}
\usepackage{cite}
\usepackage{booktabs} 
\usepackage[hidelinks]{hyperref}

\IEEEoverridecommandlockouts

\title{
Physical Limits of Proximal Tumor Detection via MAGE-A Extracellular Vesicles
}

\newcommand{\orcidiconAso}{\href{https://orcid.org/0009-0002-8206-0463}{\includegraphics[scale=0.1]{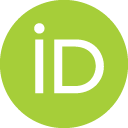}}}
\newcommand{\orcidiconOba}{\href{https://orcid.org/0000-0003-2523-3858}{\includegraphics[scale=0.1]{orcidID128.png}}}
\newcommand{\orcidiconMeb}{\href{https://orcid.org/0009-0006-3766-6684}{\includegraphics[scale=0.1]{orcidID128.png}}}

\author{
        A. Sila Okcu\orcidiconAso,~\IEEEmembership{Graduate Student Member,~IEEE,}
        M. Etem Bas\orcidiconMeb,
        and Ozgur B. Akan\orcidiconOba,~\IEEEmembership{Fellow,~IEEE}             
        \thanks{The authors are with  the Internet of Everything (IoE) Group, Electrical Engineering Division, Department of Engineering, University of Cambridge, Cambridge CB3 0FA, UK 
        (e-mails: \{aso32, meb208, oba21\}@cam.ac.uk).}
        \thanks{O. B. Akan is also with the Center for neXt-generation Communications (CXC), Department of Electrical and Electronics Engineering, Ko\c{c} University, Istanbul 34450, Turkey (e-mail: akan@ku.edu.tr).}
}

\begin{document}

\maketitle
\begin{abstract}
Early cancer detection relies on invasive tissue biopsies or liquid biopsies limited by biomarker dilution. In contrast, tumour-derived extracellular vesicles (EVs) carrying biomarkers like melanoma-associated antigen-A (MAGE-A) are highly concentrated in the peri-tumoral interstitial space, offering a promising near-field target. However, at micrometre scales, EV transport is governed by stochastic diffusion in a low-copy-number regime, increasing the risk of false negatives.
We theoretically assess the feasibility of a smart-needle sensor detecting MAGE-A–positive microvesicles near a tumour. We use a hybrid framework combining particle-based Brownian dynamics (Smoldyn) to quantify stochastic arrival and false-negative probabilities, and a reaction–diffusion PDE for mean concentration profiles. Formulating detection as a threshold-based binary hypothesis test, we find a maximum feasible detection radius of $\simeq 275\,\mu\text{m}$ for a $6000\,\text{s}$ sensing window. These results outline the physical limits of proximal EV-based detection and inform the design of minimally invasive peri-tumoral sensors.
\end{abstract}


\section{Introduction}\IEEEPARstart{T}umours actively remodel their microenvironment by secreting extracellular vesicles (EVs), specifically microvesicles (MVs) carrying tumour-associated proteins, nucleic acids, and metabolites\cite{kuldkepp2019cancer}. 
EVs, including exosomes (30--150 nm) and microvesicles (100--1000 nm), are key mediators of intercellular communication in cancer. Tumour-derived MVs participate in numerous oncogenic processes \cite{turturici2014extracellular,ortiz2021extracellular}. These EVs emerge from the tumour surface and diffuse through the interstitial space before entering lymphatic or vascular circulation. Because this peri-tumoral region lies only tens to a few hundred micrometres (typically 50–200 µm) from the tumour boundary, it contains EV concentrations orders of magnitude higher than those found in circulation. This creates a unique opportunity for proximal molecular sensing. 

Current diagnostic paradigms fail to exploit this proximal regime. Liquid biopsies, while non-invasive, suffer from extreme biomarker dilution, tumour-derived EVs are diluted by approximately $10^6–10^9\times$ upon entering milliliter-scale blood volumes, often reducing signals below detection limits\cite{heitzer2019current}. Conversely, conventional tissue biopsies sample spatially but lack molecular guidance, often missing malignant regions in heterogeneous tissues \cite{heitzer2019current}. This diagnostic gap stands in contrast to recent advances in sensing technology. Electrochemical aptasensors and microfluidic devices now report limits of detection (LOD) as low as $19-100$ particles/$\mu$L, with ultra-sensitive platforms achieving $\sim$45 particles/mL under ideal conditions \cite{gao2025electrochemical,kim2021diffusion}.
To bridge this gap, we propose a smart needle like sensor positioned directly within the peri-tumoral interstitial fluid. Conceptually aligned with the Internet of Bio-Nano Things (IoBNT) paradigm, this approach targets the molecular communication channel between the tumour and the sensor before systemic dilution occurs \cite{lopez2024immunoassays,kuscu2021internet}.

We focus on Melanoma-associated antigen A (MAGE-A), a family of cancer-testis antigens. Since MAGE-A is highly expressed in diverse malignancies but virtually absent in healthy somatic tissue, it serves as a highly specific target for such near-field detection\cite{kuldkepp2019cancer}. However, detecting these biomarkers in the interstitial space presents unique physical challenges. Unlike blood, the tumour microenvironment is not a well-mixed fluid. It is a dense, heterogeneous hydrogel (the Extracellular Matrix or ECM) that significantly hinders nanoparticle transport. This hindrance is governed by the ECM tortuosity ($\lambda$), which reduces the effective diffusion coefficient ($D_{\mathrm{eff}} = D_0/\lambda^2$) \cite{stylianopoulos2013combining}. 
Furthermore, at these micron scales, transport is not continuous but discrete. The arrival of individual vesicles is governed by rare-event Poisson statistics rather than mean concentration gradients. From a molecular communication (MC) perspective, the tumour acts as a stochastic transmitter and the sensor as an absorbing receiver \cite{aktas2024odor, akan2016fundamentals}. Standard continuum models often fail to capture the discrete fluctuations that drive false negatives in this low-copy-number regime \cite{andrews2010detailed}. Realistic modeling must also account for physical receiver constraints, such as binding kinetics and stochastic noise, as seen in FET-based biosensor models \cite{kuscu2019transmitter}. 

Therefore, we employ a hybrid framework combining Brownian dynamics with reaction-diffusion theory to quantify the physical limits of proximal tumour detection. We analyse the feasibility of detecting tumour-derived, MAGE-A bearing MVs by simulating a minimally invasive sensor placed within the peri-tumoral space. Smoldyn-based Brownian dynamics simulate discrete EV trajectories to quantify arrival distributions and false-negative probabilities, while a reaction–diffusion PDE characterizes the deterministic mean concentration field. Together, these models determine how biophysical parameters such as ECM tortuosity, secretion rate, and background EV flux govern the feasibility of spatially resolved molecular diagnostics.
\begin{figure}[t!]\centering\includegraphics[width=0.3\textwidth]{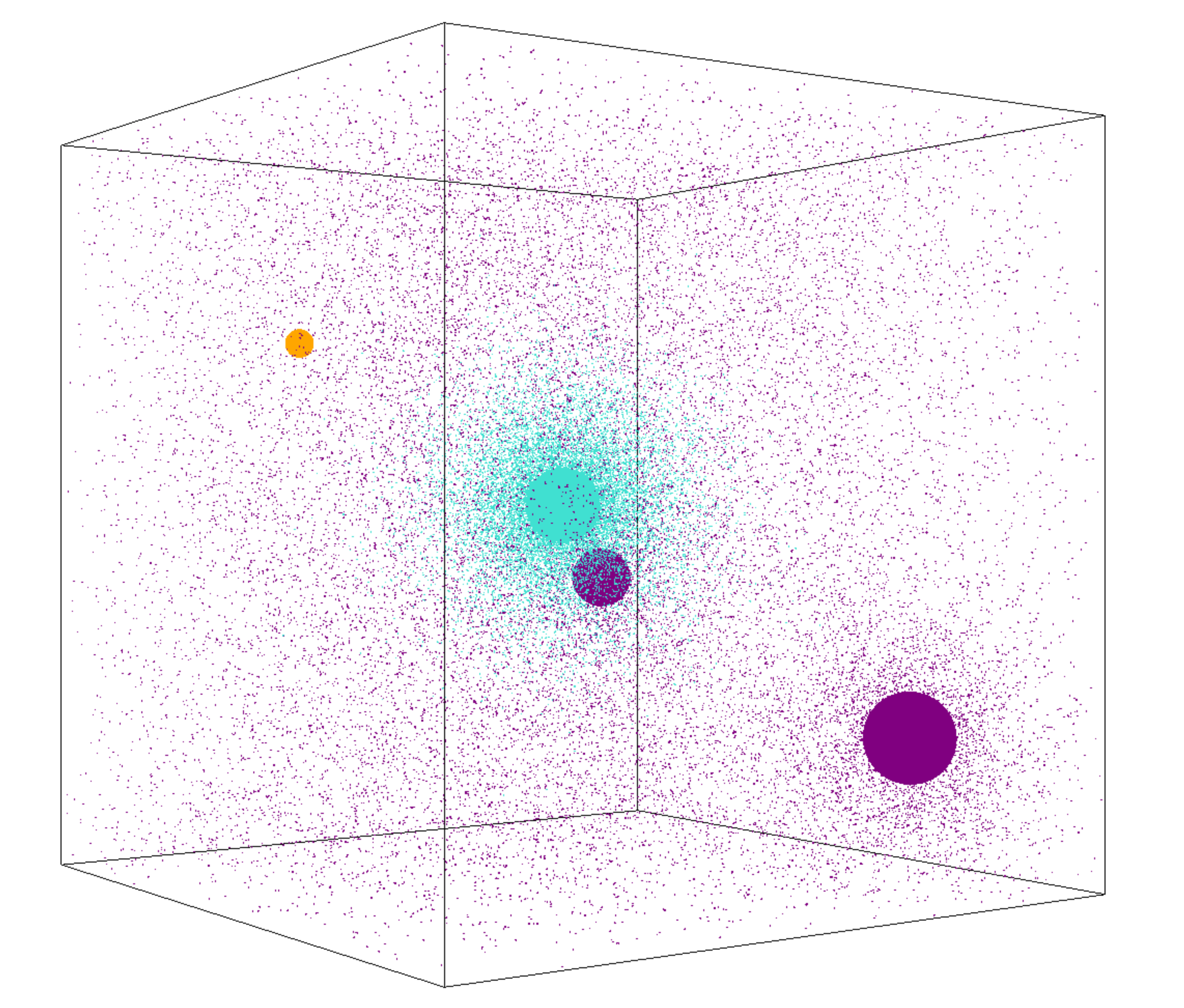}\caption{3D Smoldyn simulation showing the MAGE-A+ tumour source (turquoise), background noise sources (purple), and sensor (orange). Scattered points indicate diffusing MVs.}\label{fig:smoldyn_sim}\end{figure}

\section{Methodology}

The aim of this study is to determine whether tumour-derived MVs carrying the cancer-testis antigen MAGE-A can be detected by a microscale sensor positioned within the peri-tumoral space. We combine stochastic Brownian-dynamics simulations in Smoldyn and a continuum reaction–diffusion PDE for the mean concentration field. 
To enable comparison with in vitro biosensor LOD, we map the Smoldyn arrival counts to an effective local concentration, 
which represents the equivalent flux-derived concentration a biosensor with surface area $4\pi a_r^2$ would sample during an observation time $T$.

\subsection{System Representation}
The tumour–sensor configuration is treated as a diffusion-driven molecular communication link. The tumour is modeled as a spherical micro-lesion of radius $R_t = 50~\mu$m ($\simeq 130$cells), secreting MVs uniformly across its surface. Sensors are placed at centre–centre distances of $d\!=\!275$, $325$, and $450~\mu$m. The ECM is represented as a porous hydrogel with constant tortuosity around $\lambda\!=\!1.5$–$3.0$, consistent with literature\cite{stylianopoulos2013combining}. The sensor is modeled as a spherical absorbing receiver of radius $a_r = 20~\mu$m \cite{kuscu2016physical}. Perfect absorption represents an upper bound on performance; a capture-efficiency factor is introduced later to account for binding kinetics. The main observable is arrival count $N_T$ over $T=6000$~s window.

\subsection{Biophysical Parameterisation}

Microvesicles are treated as spherical nanoparticles whose free diffusion coefficient is given by the Stokes–Einstein relation with parameters given in Table.~\ref{tab:parameters}. This yields $D_0 \approx 0.9 \times 10^{-12}$~m$^2$/s, consistent with nanoparticle diffusion measurements in biological fluids\cite{stylianopoulos2010diffusion}.
Effective diffusivity in tissue is reduced according to
$
D_{\mathrm{eff}} = \frac{D_0}{\lambda^2},
$
where $\lambda$ is the ECM tortuosity.
A micro-lesion assuming cellular packing density of $5\times10^7$–$10^8$ cells/cm$^3$. Tumour-derived EV secretion rates vary widely, but even high-secreting immune cells produce only 24–66 EVs/min \cite{auber2022estimate}. We therefore use $q_{\text{cell}} = 20$~EV/cell/min as a generous but physiologically plausible value. This yields a total MV emission rate
$
Q_{\text{total}} \approx 40\text{–}45~\text{EV/s},
$
consistent with the micro-lesion regime.
Not all tumour-derived MVs exhibit MAGE-A on their surface. Thus, we introduce $p_{\text{tag}}$ as the fraction of MVs carrying accessible MAGE-A. Literature reports vary but consistently show partial expression; therefore we use $p_{\text{tag}} = 0.2$ in the corrected detection model\cite{kuldkepp2019cancer}.
Interstitial clearance is modeled as a first-order decay process with half-life $t_{1/2} = 2$~h, giving
$
k_{\text{clear}} = \ln(2)/t_{1/2} \approx 10^{-4}\,\text{s}^{-1}.
$
All parameters are applied consistently across Smoldyn simulations and PDE calculations.

\begin{table}[htbp]
\centering
\caption{Biophysical parameters used in the molecular communication model.}
\label{tab:parameters}
\begin{tabular}{|l|c|l|}
\hline
\textbf{Parameter} & \textbf{Symbol} & \textbf{Value} \\
\hline
Body temperature & $T$ & 310\,K \\
Interstitial viscosity & $\eta$ & 2.0\,mPa$\cdot$s \\
MV radius & $r_p$ & 250\,nm \\
Free diffusion coefficient & $D_0$ & $0.9\times10^{-12}$ m$^2$/s \\
Tumour radius & $R_t$ & 50~$\mu$m \\
Cells per tumour & $N_{\text{cell}}$ & $\sim130$ \\
Secretion rate per cell & $q_{\text{cell}}$ & 20 EV/min \\
Tumour EV emission & $Q_{\text{total}}$ & 40--45 EV/s \\
Tagged MV fraction & $p_{\text{tag}}$ & 0.2 \\
MV half-life & $t_{1/2}$ & 2 h \\
Clearance rate & $k_{\text{clear}}$ & $10^{-4}$ s$^{-1}$ \\
\hline
\end{tabular}
\end{table}

\subsection{Smoldyn Stochastic Simulation}

Vesicle motion is simulated using overdamped Langevin dynamics,
\[
\mathbf{x}(t+\Delta t) = \mathbf{x}(t) + \sqrt{2D_{\text{eff}}\Delta t}\,\boldsymbol{\xi},
\]
with timestep $\Delta t=1$~s, yielding RMS displacements $\ll a_r$ and numerically stable capture. Vesicles are injected at the tumour boundary following a Poisson process of rate $Q$. Clearance occurs with probability $1 - e^{-k_{\text{clear}}\Delta t}$. Simulations are performed in a $1000^3~\mu$m reflective domain; the diffusion length $\sqrt{2D_{\text{eff}}T}\approx100~\mu$m ensures negligible boundary influence.
For each parameter set, $M=100$ realisations provide empirical means, variances, and false-negative probabilities.
The primary output of each run is the integer-valued arrival count $N_T$, defined as the number of vesicles that contacted and are absorbed by the receiver surface during the observation window $T=6000$\,s. The simulation duration was chosen as it since it is comparable to the MV half-life ($t_{1/2} = 2$ h), which ensures that clearance dynamics are captured. The ensemble of arrival counts yields an empirical probability mass function, a mean arrival rate, a variance, and a probability of zero arrivals. These metrics quantify the degree of stochasticity inherent to the communication channel and provide the basis for hypothesis testing.
In Fig.~\ref{fig:smoldyn_sim}, non-MAGE-A vesicles represent background EVs emitted by healthy tissue. These particles are included in the false-positive model but do not affect Smoldyn arrival counts for the tumour-only simulations, whose role is formalized in the detection logic. 

\subsection{Continuum Reaction–Diffusion PDE}

\begin{figure*}[t!]
    \centering
    \includegraphics[width=1\textwidth]{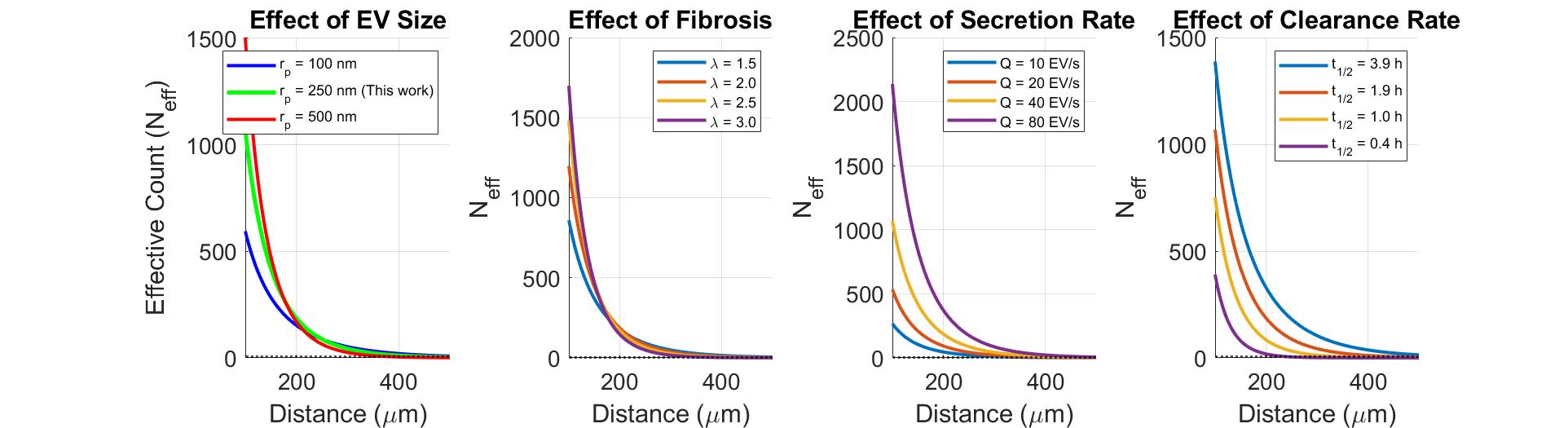}
    \caption{Sensitivity to EV biophysical properties and microenvironment.}
    \label{fig:sweep}
\end{figure*}

\begin{figure}[t!]
    \centering
    \includegraphics[width=0.5\textwidth,height=5cm]{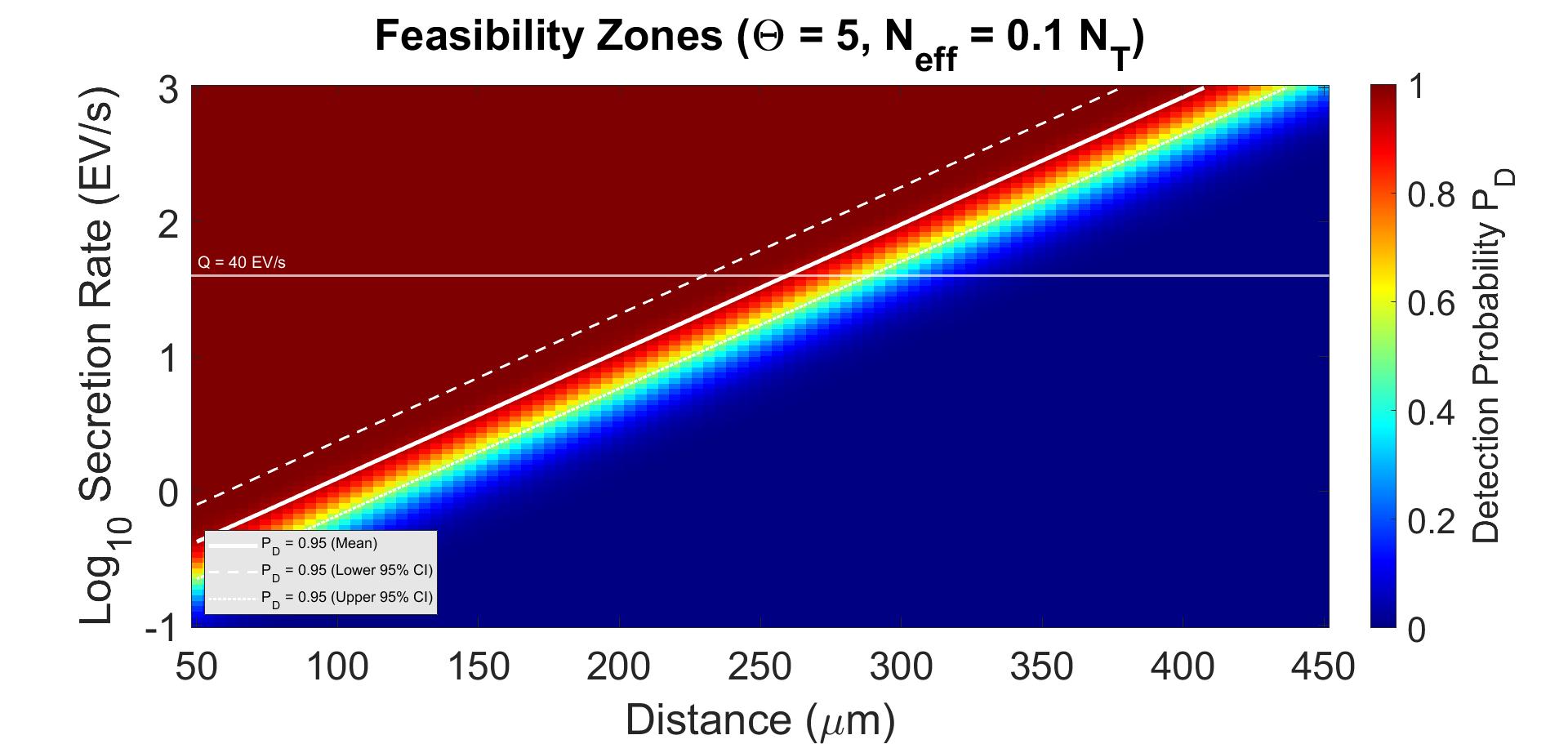}
    \caption{Operating regime map showing 95\% detection contour.}
    \label{fig:heat_meap}
\end{figure}

To obtain the mean concentration field, we solve
\begin{equation}
\frac{\partial C}{\partial t}
=
\frac{1}{r^2}
\frac{\partial}{\partial r}
\left(D_{\mathrm{eff}} r^2 \frac{\partial C}{\partial r}\right)
-
k_{\mathrm{clear}} C, \quad r \ge R_t,
\end{equation}
with boundary conditions
\[
-D_{\mathrm{eff}}{\partial C}/{\partial r}|_{r=R_t} = J_{\mathrm{EV}}, 
\qquad
C(r_{\max}) = 0,
\]
where $r_{\max} = R_t + 450~\mu$m.

The assumption of spherical symmetry is appropriate for micro-lesions whose curvature dominates over macroscopic anisotropies, and for sensors placed at fixed radial distances.

The PDE is solved using 200 spatial nodes over 6000~s. Despite the reflective–Dirichlet mismatch noted above, the PDE and Smoldyn means agree well within the region of interest.
The steady-state solution,
\[
C(r) = \frac{J_{\mathrm{EV}} R_t^2}{D_{\mathrm{eff}} r}
\exp\!\left[-(r-R_t)\sqrt{\frac{k_{\mathrm{clear}}}{D_{\mathrm{eff}}}}\right],
\]
predicts the mean concentration. The expected receptor flux scales with;
$
E[N_T] \propto 4\pi a_r D_{\mathrm{eff}} C(r_{\mathrm{Rx}})\, T,
$
linking continuum fields to discrete arrival counts.
\begin{figure*}[t]
\centering
\makebox[0.82\textwidth][c]{%
\begin{minipage}{0.82\textwidth}
\centering

\subfloat[]{%
  \includegraphics[width=0.45\linewidth]{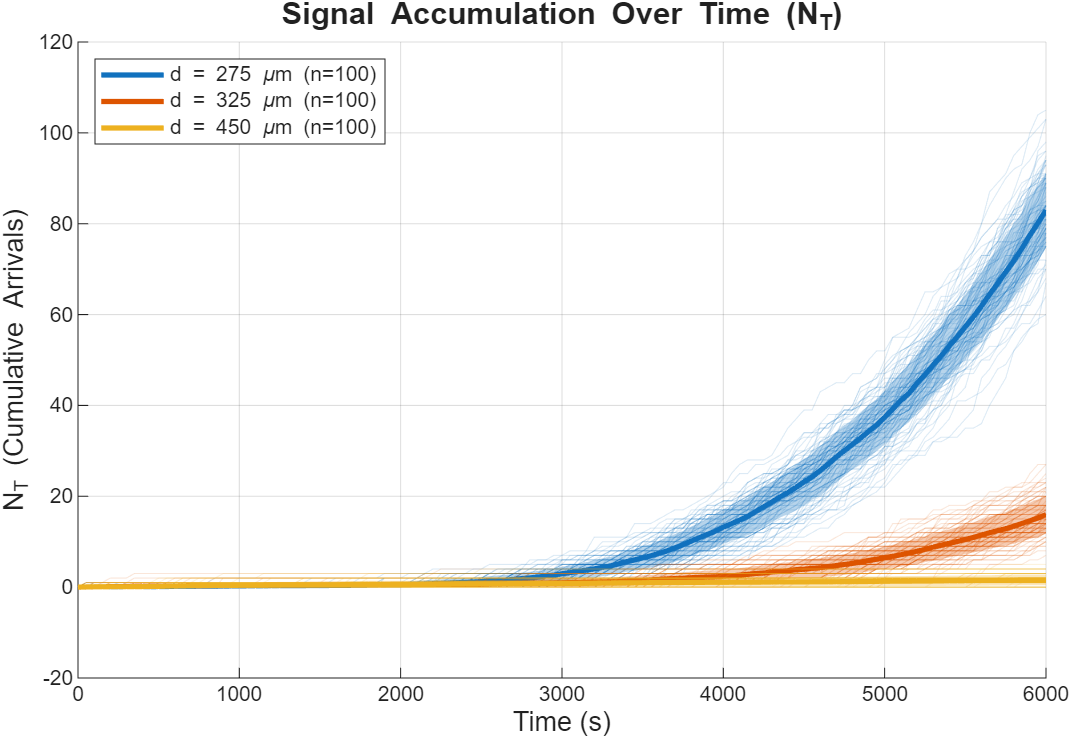}%
  \label{fig:transient}
}\hspace{1em}
\subfloat[]{%
  \includegraphics[width=0.45\linewidth]{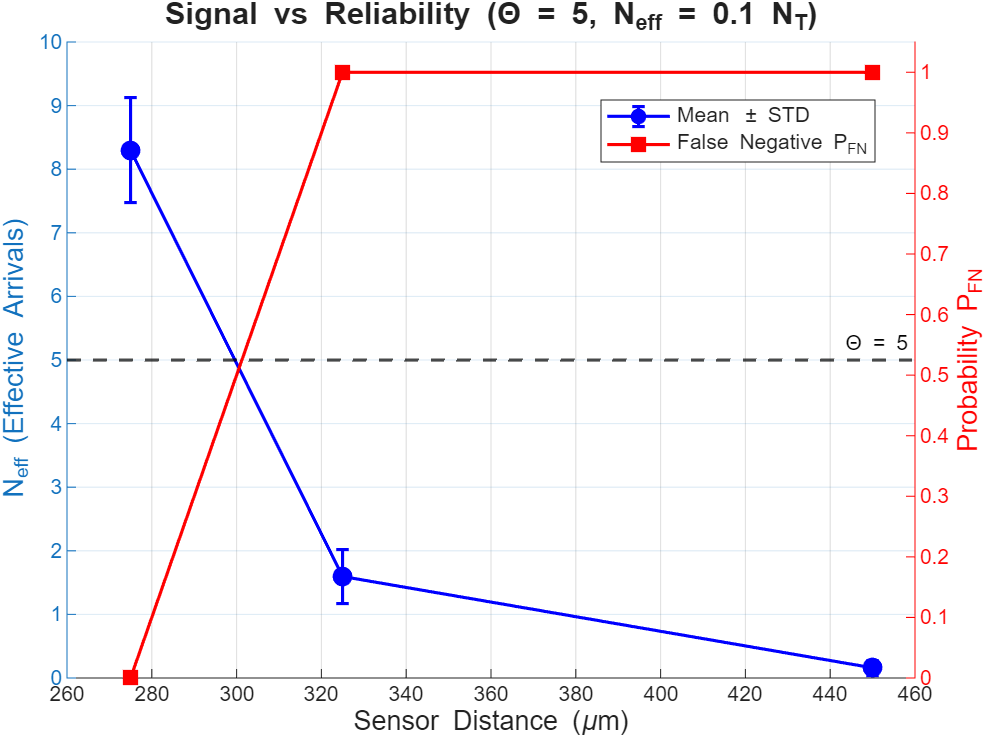}%
  \label{fig:pfn}
}

\vspace{0.6em}

\subfloat[]{%
  \includegraphics[width=0.45\linewidth]{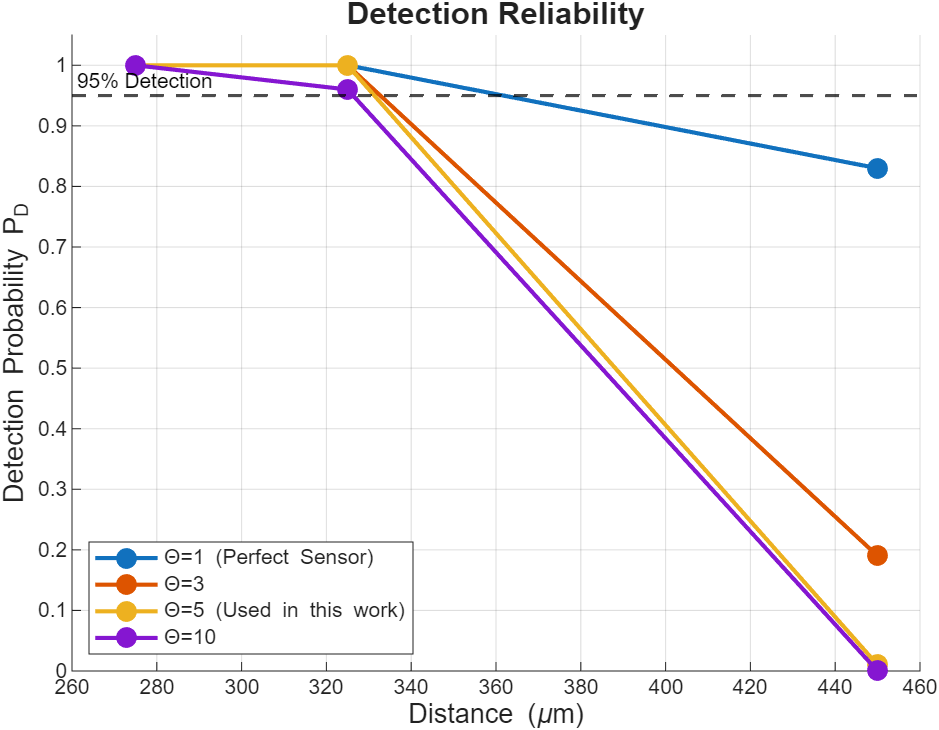}%
  \label{fig:detection}
}\hspace{1em}
\subfloat[]{%
  \includegraphics[width=0.45\linewidth]{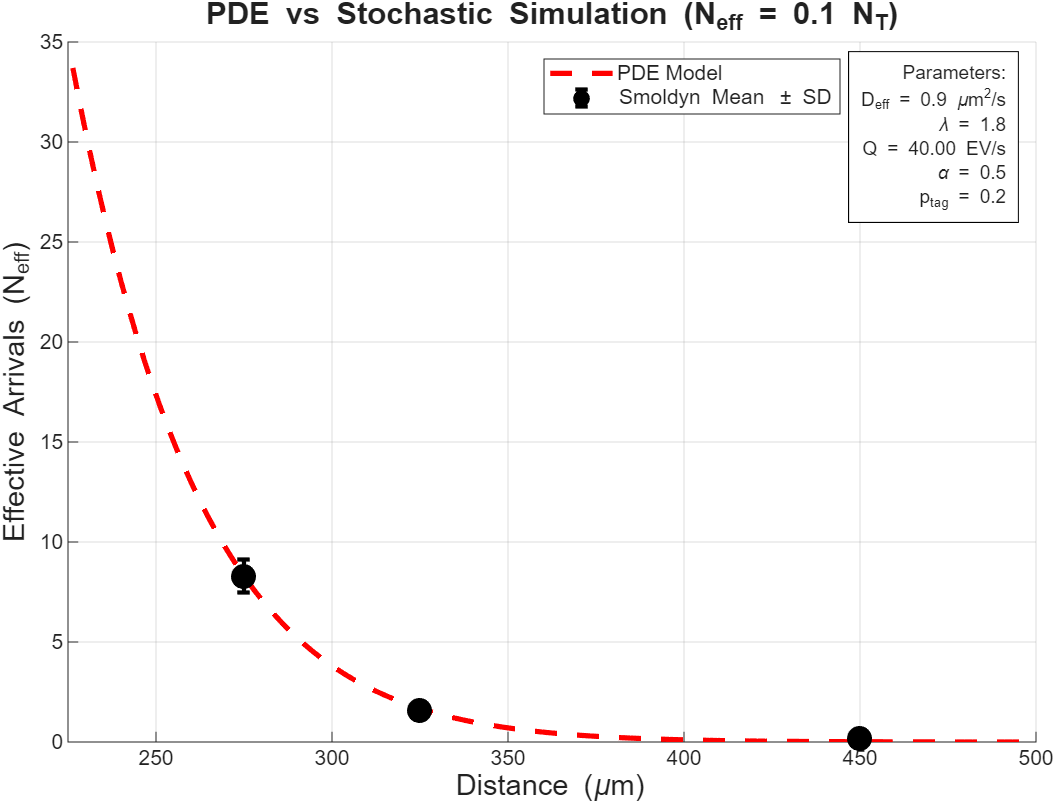}%
  \label{fig:pde}
}

\end{minipage}%
}
\caption{Computational modeling of EV transport and detection.
(a) Cumulative arrival kinetics at different sensor distances from tumour center ($N_T$).
(b) Trade-off between signal magnitude and false negative probability.
(c) Detection probability as a function of distance for multiple thresholds.
(d) Validation of mean-field PDE against stochastic Smoldyn simulations.}
\label{fig:combined}
\end{figure*}

\subsection{Feasible Detection Radius}
In Fig.\ref{fig:heat_meap}, we define the Maximum Feasible Radius as the distance at which the False Negative Probability ($P_{\mathrm{FN}}$) remains below 5\%.These results suggest that reliable detection requires either high secretion rates or short sensor-tumour distances. At $Q$ = 40 MV/s, detection remains clinically reliable up to $\approx 275-280 ~\mu$m from the center of the tumour.

\subsection{Different Parameters}
Fig.~\ref{fig:detection} illustrates the trade-off between sensitivity and specificity across different detection thresholds. We selected $\Theta = 5$ to maintain a robust margin above the expected background noise ($\mathbb{E}[N_{\text{bg}}] \approx 3$) while ensuring high detection probability at clinically relevant distances. The sensitivity analysis, Fig.~\ref{fig:sweep}, confirms that signal strength scales proportionally with secretion rate ($Q$) and inversely with clearance rate ($k_{\mathrm{clear}}$). Increased ECM tortuosity associated with fibrosis ($\lambda > 2.0$) significantly dampens diffusive transport, reducing the effective detection radius compared to healthy tissue, while smaller EV radii ($r_p$) enhance range via higher diffusivity.

\subsection{Detection Logic}

Under the null hypothesis $H_0$, the sensor is located in healthy tissue and no MAGE-A positive vesicles are emitted. However, healthy cells also emit EVs, and the sensor may capture particles non-specifically. We model this background noise as Poisson process with rate $\lambda_{\text{noise}}$.
Since MAGE-A is a cancer-testis antigen whose expression in normal adult tissues is restricted to germ cells and trophoblasts, the probability of MAGE-A positive EVs appearing in healthy tissue is extremely low\cite{kuldkepp2019cancer}. We estimate $\lambda_{\text{noise}} \approx 5 \times 10^{-4}$ s$^{-1}$, yielding an expected background count of $\mathbb{E}[N_{\text{bg}}] \approx 3$ over $6000s$.
Under the alternative hypothesis $H_1$, the sensor is near a tumour and the vesicles arrive according to the stochastic process described above. A threshold detector declares the presence of tumour whenever the arrival count exceeds a cutoff value $\Theta$. 
To account for realistic sensor limitations, we introduce two correction factors. First, not all tumour-derived EVs express MAGE-A on their surface; we define $p_{\text{tag}}$ as the fraction of tumour MVs carrying the antigen. Second, real sensors have finite binding kinetics and receptor saturation rather than perfect absorption; we define $\alpha$ as the effective capture efficiency. The detectable signal is therefore modelled as
$
N_{\text{eff}} = \alpha \cdot p_{\text{tag}} \cdot N_T,
$
where $N_T$ is the number of EVs contacting the sensor surface. For our analysis, we use $p_{\text{tag}} = 0.2$ and $\alpha = 0.5$, which yields $N_{\text{eff}} = 0.1 \times N_T$. Our reported results using $N_T$ directly represent an upper-bound estimate of detection performance. ROC analysis confirms that this value lies near the optimal trade-off between false positives and false negatives.

\section{Results and Analysis}

The objective of the analysis is to determine whether tumour-derived MAGE-A–positive MVs can be reliably detected and the detection interval by a microscale absorber positioned within the peri-tumoral space. The results combine Smoldyn-derived stochastic arrival statistics and continuum PDE-derived mean concentration fields.
Importantly, tumour EVs exist alongside large populations of \emph{non-MAGE-A} EVs released by stromal and immune cells; modeling this background is also essential for understanding false-positive behaviour in proximal sensing. In our study, the tumour-derived MAGE-A-positive vesicles constitute the signal, while non-MAGE-A EVs represent a potential noise source when mapping arrival counts to sensor-level detection thresholds.

\subsection{Stochastic Signal Strength and Variance}

\begin{figure*}[t!]
    \centering
    \includegraphics[width=\textwidth, height=5cm]{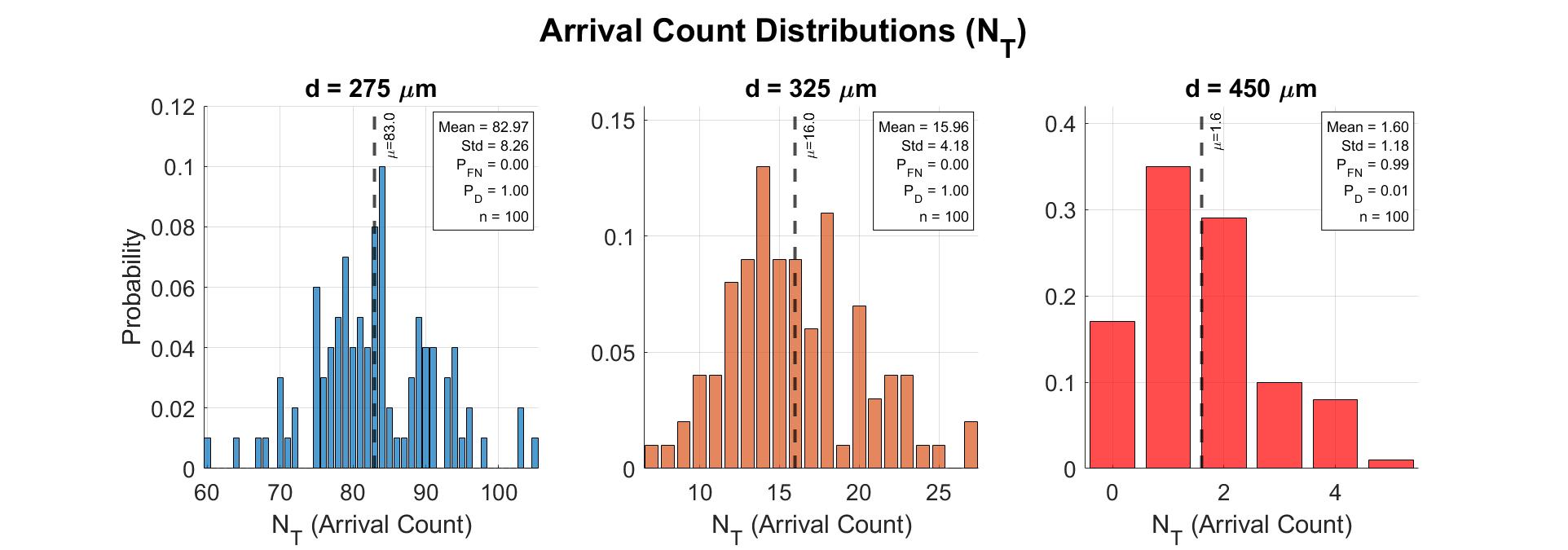}
    \caption{Histograms of arrival count distributions (raw $N_{T}$)}
    \label{fig:histogram}
\end{figure*}

Fig. ~\ref{fig:transient} shows the distribution of EV arrival counts after $N = 100$ runs. At 275 $\mu$m distance from the tumour centre, the sensor receives a mean raw signal of $\mu = 83.0$ MVs, corresponding to $N_{\text{eff}} = 8.3$ MVs ($\sigma = 0.82$), with 0\% false negatives ($P_{FN} = 0$). At 450 $\mu$m, the mean falls to $N_T = 1.6$ ($N_{\text{eff}} = 0.16$, $\sigma = 0.12$), resulting in complete detection failure ($P_{FN} = 1.00$). 
Fig. \ref{fig:pfn} shows the inverse relationship between signal strength and detection reliability. The mean arrivals decay nearly exponentially with distance, while the false negative probability remains low until approximately 275 $\mu$m. Beyond this point, $P_{FN}$ increases sharply, defining the maximum clinically feasible detection range for the given parameters. The arrival count distributions shown in Fig. \ref{fig:histogram} clearly reveal the stochastic nature of EV-based detection.

\subsection{Validation against Continuum Theory \& Comparison with Sensor Detection Limits}The stochastic mean arrival counts were compared against the deterministic concentration profile derived from the reaction–diffusion PDE (Fig.~\ref{fig:pde}). The Smoldyn results closely track the analytical exponential decay predicted by the continuum model, confirming that the stochastic variance observed is a genuine physical property of the low-copy-number channel and not a simulation artifact.
To compare with real sensor specifications, we estimate local concentration from the diffusive flux at an absorbing sphere ($C_{\text{local}} = \frac{N_T}{4\pi a_r D_{\text{eff}} T}$). 
At 275 $\mu$m ($N_T = 83$), this yields $\approx
6 \times 10^4$ particles/$\mu$L, far exceeding the reported LOD of state-of-the-art aptasensors (19--100 particles/$\mu$L). At 450 $\mu$m ($N_T = 1.6$), the concentration drops to $\approx 1.2 \times 10^3$ particles/$\mu$L, still well above typical LODs. This confirms that detection failure at large distances is primarily due to stochastic variability rather than sensor sensitivity limits.

\subsection{Limitations \& Future Work}
Our model relies on several simplifications. We assume a homogeneous ECM without interstitial convection. The idealized spherical micro-lesion neglects irregular tumour shapes and spatial heterogeneity.
The receiver is modeled as a perfect absorber with scalar efficiency corrections. This approximates, but does not fully capture, complex binding kinetics and receptor saturation. Finally, simulations were limited to transient timescales; steady-state analysis was constrained by computational cost of 3D Brownian dynamics.
Future work will incorporate convection-diffusion dynamics and extend this framework to multi-sensor configurations. By leveraging Bayesian inference or learning-based methods, spatially distributed measurements could be used to localize the tumour source\cite{okcu2025smell}. We also aim to validate the detection threshold against experimental noise floors of aptasensors.

\section{Conclusion}
This paper presents a theoretical framework for validating the feasibility of proximal tumour detection using microvesicles as biomarkers. By bridging tumor biology with transport physics, we identified a critical detection window within the interstitial space. Our results demonstrate that for a standard micro-lesion with a secretion rate of $\approx 40$ EV/s, a sensor can achieve reliable detection up to a maximum radial distance of $275~\mu m$. Beyond this range, stochastic arrival fluctuations cause the false negative rate to rise sharply, rendering detection unreliable regardless of sensor sensitivity. This defines the physical limit for minimally invasive smart-needles, suggesting they must be positioned within $\approx 225~\mu m$ of the tumour boundary to beeffective.

\bibliographystyle{IEEEtran}
\bibliography{references}

\end{document}